\documentclass[12pt,twoside]{article}
\usepackage{inputenc}
\usepackage[english,russian]{babel} 
\usepackage{jcem}
\usepackage{amsfonts}
\usepackage{amsmath}
\usepackage{amssymb}
\usepackage{amsthm}
\usepackage{graphicx}
\usepackage{graphicx, caption}
\usepackage{epsfig, epsf}
\usepackage{epstopdf}
\begin{document}
\def\Re{\mathop{\rm Re}\,}
\def\B{\mathop{\frak B}}
\def\R{\mathop{\Bbb R}}
\def\A{\mathop{\frak A}}
\def\G{\mathop{\bf G}}
\def\U{\mathop{\frak U}}
\def\H{\mathop{\cal H}}
\def\Im{\mathop{\rm Im}\,}
\def\dom{\mathop{\rm dom}\,}
\def\dist{\mathop{\rm dist}}
\def\grad{\mathop{\rm grad}}
\renewcommand{\proof}{\vspace{2mm}\hspace{-7mm}\textit{Proof.}}
\renewcommand{\endproof}{\begin{flushright} \vspace{-2mm}$\Box$\vspace{-4mm}
\end{flushright}}
\newcommand{\phan}{\hspace*{0cm}}
\newcommand{\comment}{}

\begin{flushleft}
\textbf{MSC 74S20, 65Z05, 74A15} 
\end{flushleft}
\author{M.V. Polyakov, {\rm Volgograd State University, Volgograd, Russia, maxim.v.polyakov@gmail.com},\\
\ A.V. Khoperskov, {\rm Volgograd State University, Volgograd, Russia, khoperskov@volsu.ru},\\
\ A.V. Svetlov, {\rm Volgograd State University, Volgograd, Russia, andrew.svetlov@volsu.ru}
}
\title{SIMULATION MODELING OF RADIATION FIELD IN BIOLOGICAL TISSUE OF MAMMARY GLAND}
\maketitle{}

\begin{abstract} \begin{tabular}{p{0mm}p{139mm}}
&\noindent {\footnotesize \qquad
In this paper we develop a mathematical model of the distribution of microwave electric field in heterogenous biological tissue of mammary gland. We use this model to investigate the efficiency of the medical diagnostic method based on microwave thermometry. Also we run a numerical modeling of electromagnetic field in biotissue of mammary gland for various sets of the spatial structure of biotissue. The small-scale structure is caused by a complex combination of several components: blood flows, fat tissue, muscle tissue, milk lobules, skin. Next we vary in the model the spatial structure of the tissue to evaluate the effect of the heterogeneous structure of the tissue on the distribution of the electromagnetic field in the volume of mammary gland.

\qquad}
\end{tabular}\end{abstract}

\markboth{M.~V. Polyakov, A.~V. Khoperskov, A.~V. Svetlov}{}

\section*{Introduction}
\hspace{0.7 cm}

Among the various methods for cancer detection, radiothermometry is extremely important, because it allows rapid mass screening \cite{Barett,Akki-Arunachalam-2014,losev,novochadov}. The method is based on the experimental measurement of the temperature inside the biological tissue at various points. We will discuss the results of a series of simulation modeling of radiation fields in the microwave range for the agenda to increase the efficiency of radio thermometry method of cancer detection \cite{vesnin}.

One of the problems is the variability of measurements of the temperature field and individual physiological characteristics of biological tissue in different people \cite{Carr,Foster}.
We base our model on numerical integration of the Maxwell equations.
The complex spatial structure of the tissue and its heterogeneity on small scales requires the use of unstructured numerical grids to calculate the electric field \cite{polyakov}.

We compare the distribution of the radiation field $\vec {E}(\vec {r}) $ and its power $P_d$ for different models against each other.

\section{Basic models}
\subsection{Formulation of the problem}
\hspace{0.7 cm}

We consider a model of the mammary gland in the form of a hemisphere with an adjacent cylinder (fig. 1). The main internal components are muscle and fat tissue, breast lobules, blood, and skin.
\begin{figure}[!h]
\centering{
  \includegraphics[width=6cm]{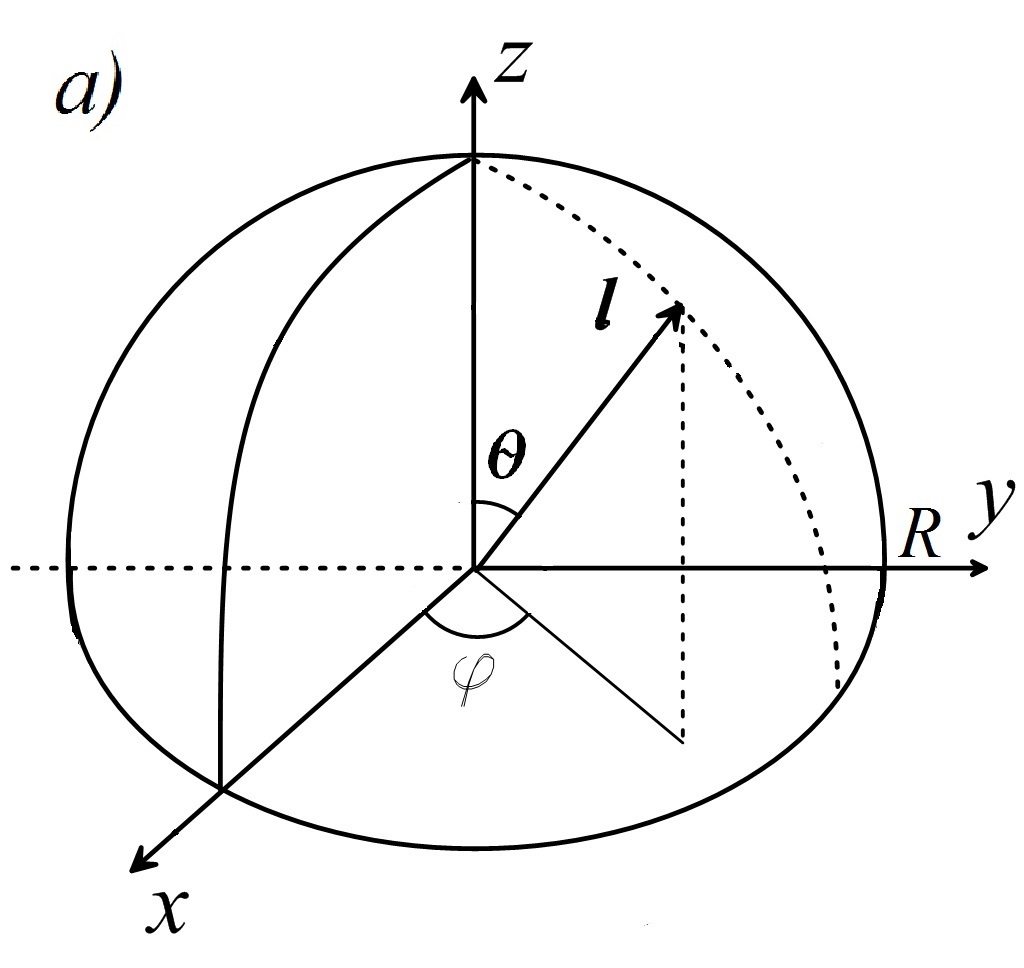}
  \includegraphics[width=4.2cm]{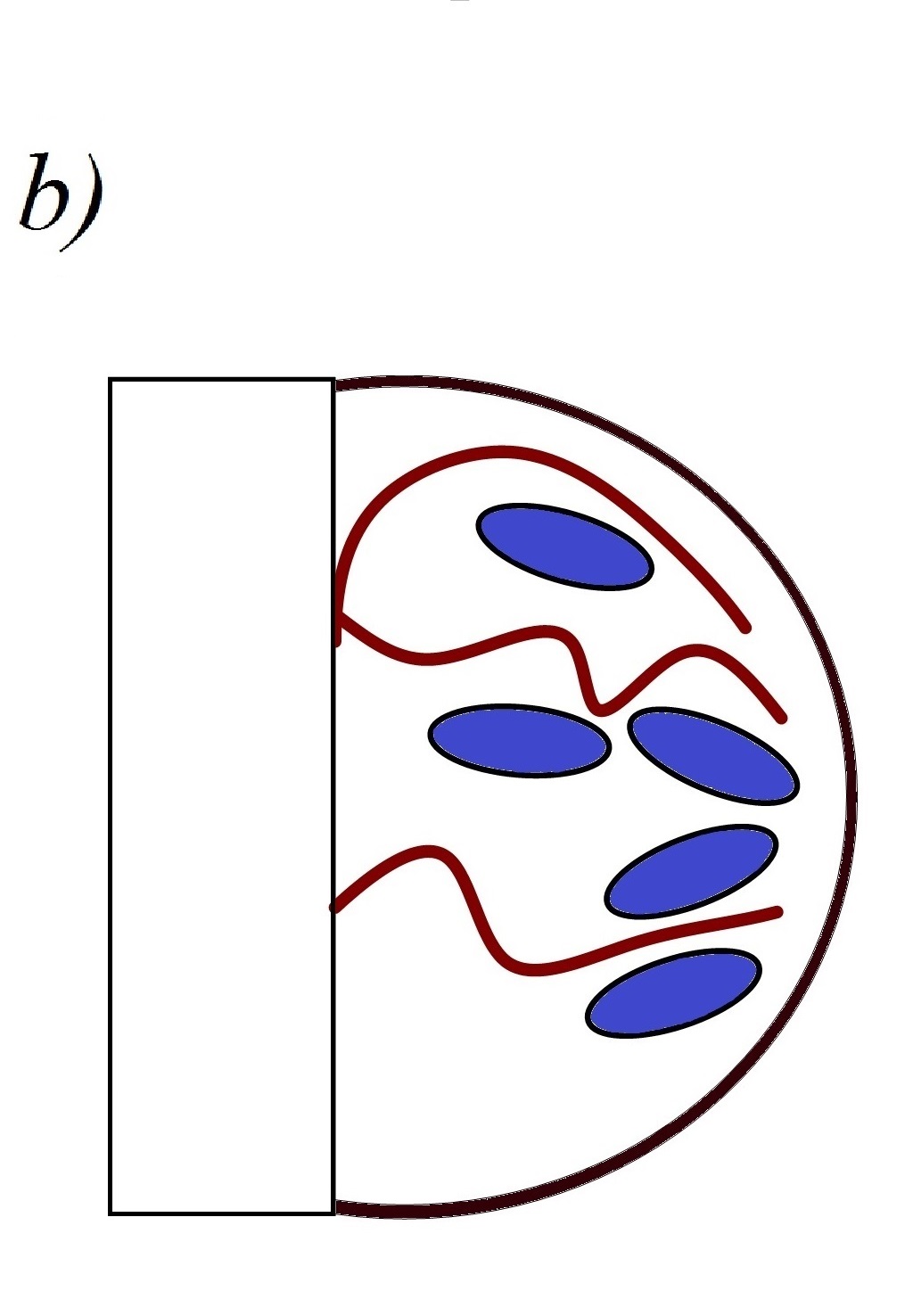}
  \includegraphics[width=4.2cm]{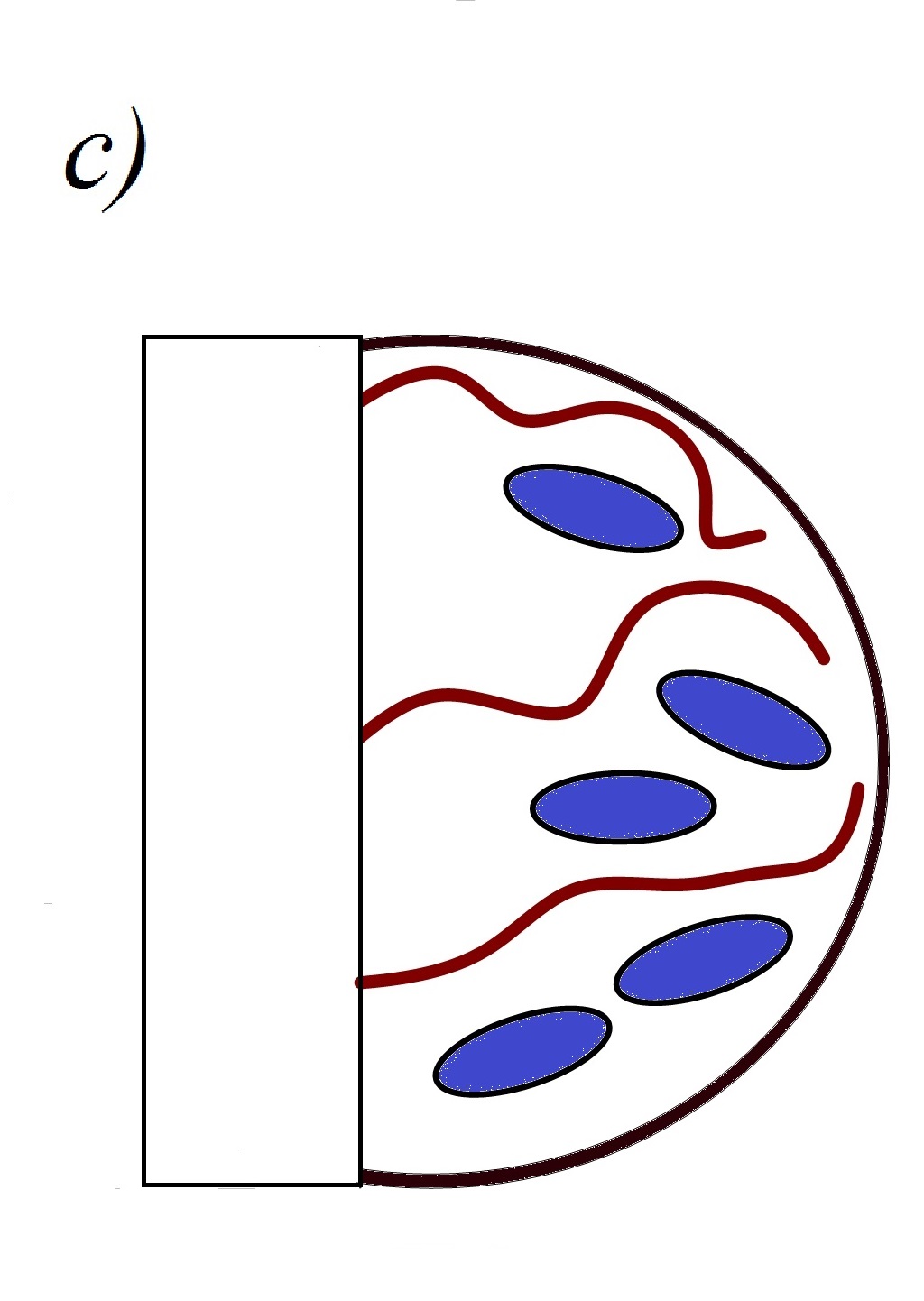}
    \renewcommand{\figurename}{Fig.}
  \caption{\small The geometry of the breast measurement model (a), Schematic representation of two different internal structures of the tissue (b, c)}
  }
\end{figure}

There is a natural variation of this internal structure in women. Individual differences can be significant, and this leads to differences in spatial distributions of physical parameters of biological tissues.
 We use a microwave antenna in the range 1--1.5 GHz to measure the internal temperature. Formation of electromagnetic field inside the tissue depends on its conductivity $\sigma$, permittivity $\varepsilon$ and resistivity ${\cal R}$. Various biological components have different values of $\sigma$, $\varepsilon$, ${\cal R}$. We build sets of models with different spatial structure within the limits of natural variability and we calculate the electric field for each model. The simulation modeling allows us to estimate an error of internal temperature measurement by the method of microwave thermometry \cite{Gonzalez,losev}.

Each experiment in the series is different in number and location of the main components (milk lobules, connective and fat tissues, blood flows). The radius of the breast is the same in all models, the antenna parameters are fixed. Each component has the same set of characteristics $\sigma$, $\varepsilon$, ${\cal R}$ (Table 1).

We fix the number of breast lobules and total length of the blood streams, and we vary only the spatial location.

\newpage
\subsection{Small-scale internal structure of biotissue}
\hspace{0.7 cm}

Here we consider two series of numerical experiments: 1) the antenna is located in the nipple area (let's call it the point ``0''); 2) the antenna is dislocated from the point ``0'' by $45^{\circ}$ (fig. 1).

We restrict ourselves to the set of models, where among other structural components just milk lobules ($mg$) and blood streams ($bl$) can vary their position. They are surrounded by connective and fat tissue ($mus$). Skin parameters are fixed ($sk$). The relative volumes of all four components $mg$, $bl$, $mus$, $sk$ in all models are the same:
$$
\gamma_{mg}=\frac{V_{mg}}{V_{0}}=\frac{402~\text{cm}^{3}}{1436~\text{cm}^{3}}\approx 0.28,
\gamma_{sk}=\frac{V_{sk}}{V_{0}}=\frac{143~\text{cm}^{3}}{1436~\text{cm}^{3}}\approx 0.1,
$$
 \begin{equation}\label{tissue}
 \gamma_{mus}=\frac{V_{mus}}{V_{0}}=\frac{877~\text{cm}^{3}}{1436~\text{cm}^{3}}\approx 0.61,
\gamma_{bl}=\frac{V_{bl}}{V_{0}}=\frac{14~\text{cm}^{3}}{1436~\text{cm}^{3}}\approx 0.01,
\end{equation}
where $\gamma_{mg} + \gamma_{mus} + \gamma_{sk} + \gamma_{bl} = 1$.

  We calculate the electric fields $E (x, y, z)$ for 9 different models. They depend on the spatial distributions of the permittivity $\varepsilon (x,y,z)$, resistivity ${\cal R}(x,y,z)$ and conductivity $\sigma(x,y,z)$ (Table 1). For $i$-th model ($i$ = 1, ..., 9) we have the power density $P_d =\sigma E^{2}/2$. Coefficients

 \begin{equation}\label{koef}
\beta_{ij}=\sqrt{\int\left[\frac{\sigma_{j}E^{2}_{j}-\sigma_{i}E^{2}_{i}}{\sigma_{j}E^{2}_{j}}\right]^2}dV
\end{equation}
 reflect the influence of the internal structure (components position relative to each other) on the  power density distribution of the electric field and the internal temperature.

\begin{figure}[!h]
\centering{
  \includegraphics[width=10cm]{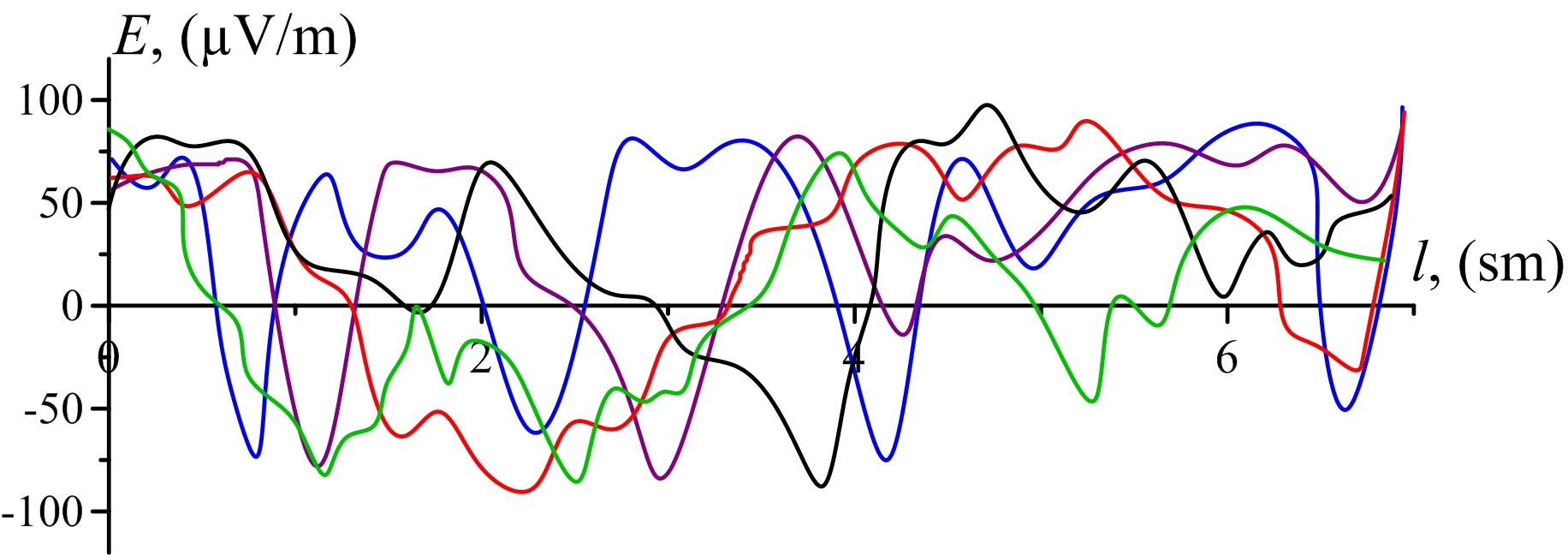}
   \includegraphics[width=10cm]{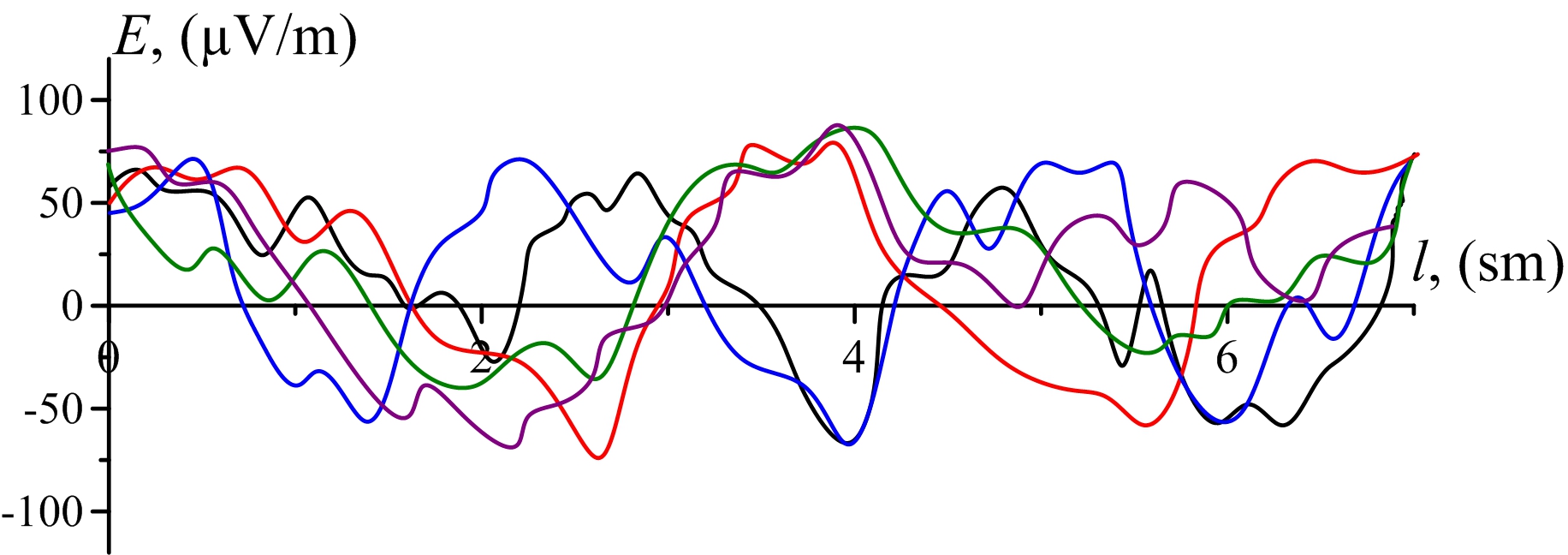}
   \renewcommand{\figurename}{Fig.}
  \caption{\small Values of $E$ along the line $l$ for two different models}
  }
\end{figure}

\begin{table}[!h]
\centering
\renewcommand{\tablename}{Table.}
\caption{Physical parameters of components}
\label{parameters}
\begin{tabular}{|l|c|c|c|c|}
\hline
                              & Skin & \begin{tabular}[c]{@{}c@{}} Mammary \\ gland \end{tabular} & \begin{tabular}[c]{@{}c@{}}Connective \\         tissue\end{tabular} & Bloodstream \\ \hline
\begin{tabular}[c]{@{}c@{}} Dielectric \\ permeability, $\varepsilon$ \end{tabular} & 55.4 & 5.5             & 46                                                                      & 1.87     \\ \hline
\begin{tabular}[c]{@{}c@{}}Electric \\ conduction, \\ $\sigma$ (1$/$Ohm) \end{tabular}           & 1.08 & 0.06            & 2.55                                                                    & 48       \\ \hline
\begin{tabular}[c]{@{}c@{}} Resistivity,\\ $\cal R$ (Ohm$\cdot$m)\end{tabular}       & 55   & 15              & 1.6                                                                     & 1.5      \\ \hline
\end{tabular}
\end{table}

\subsection{Mathematical and numerical models}

Antenna with a frequency of a few GHz allows to measure thermal radiation from biological tissues in the frequency range $\omega _{\min}\le\omega\le\omega_{\max}$ \cite{vesnin,Bardati}. The biological tissue has an inhomogeneous temperature distribution, that's why the method gives us just the weighted average temperature of some inner region $T_b^{(exp)}$. The error in the RTM method is also caused by the noise temperature of the receiver $T_{REC}$, the effects of the antenna mismatch ($S_{11}(\omega)$ coefficient), and environmental influence ($T_{EMI}$). As the result, we have to determinate the brightness temperature by the integral representation:
\begin{equation}\label{eq-TB}
T_B^{(exp)} = \int\limits_{\omega_{\min}}^{\omega_{\max}} \left\{ s_{11} \left[ T_{EMI} + \int_{V_0} W(x,y,z;\omega)\, T(x,y,z)\,dV \right] + |S_{11}(\omega)|^2 T_{REC}
\right\}\,d\omega \,,
\end{equation}
where
$s_{11} = 1 - |S_{11}|^2$ takes into account the antenna mismatch,
\begin{equation}\label{1}
W=\frac{P_d(x,y,z;\omega)}{\int_{V_b} P_d \,dV}
\end{equation}
is weight function with normalization,
\begin{equation}\label{2}
\int_{V_0} W\,dV=1,
\end{equation}
\begin{equation}\label{3}
P_d = \frac{1}{2}\sigma(x,y,z;\omega)\cdot |{\vec E(x,y,z;\omega)}|^2
\end{equation}
is power density, and $\sigma$ is electric conduction.

To construct a stationary electric field distribution, it is convenient to solve the time-dependent Maxwell equations and as the result to obtain the stationary-state:
\begin{equation}\label{eq-Makswell}
    \frac{\partial \vec{B}}{\partial t} + rot(\vec{E}) = 0 \,, \quad \frac{\partial \vec{D}}{\partial t} - rot(\vec{H}) = 0 \,,\quad \vec{B}=\mu\vec{H}\,,\quad \vec{D}=\varepsilon\vec{E} \,.
\end{equation}
It is convenient to use a uniform grid numerical scheme $x_{i+1}=x_i+\Delta{x}$, $y_{j+1}=x_j+\Delta{y}$, $z_{k+1}=x_k+\Delta{z}$:
\begin{equation}\label{eq-Maxwell-num1}
    \frac{B_x^{n+1/2}(x_i,y_{j+1/2},z_{k+1/2}) - B_x^{n-1/2}(x_i,y_{j+1/2},z_{k+1/2})}{\Delta{t}} =
\end{equation}
$$
= \frac{E_y^{n}(x_i,y_{j+1/2},z_{k+1}) - E_y^{n}(x_i,y_{j+1/2},z_{k})}{\Delta{z}} -
$$
$$
\frac{E_z^{n}(x_i,y_{j+1},z_{k+1/2}) - E_z^{n}(x_i,y_{j},z_{k+1/2})}{\Delta{y}} \,,
$$
\begin{equation}\label{eq-Maxwell-num2}
    \frac{D_x^{n}(x_{i+1/2},y_{j},z_{k}) - D_x^{n-1}(x_{i+1/2},y_{j},z_{k})}{\Delta{t}} =
\end{equation}
$$
= \frac{H_z^{n-1/2}(x_{i+1/2},y_{j+1/2},z_{k}) - H_z^{n-1/2}(x_{i+1/2},y_{j-1/2},z_{k})}{\Delta{y}} -
$$
$$
- \frac{H_y^{n-1/2}(x_{i+1/2},y_{j},z_{k+1/2}) - H_y^{n-1/2}(x_{i+1/2},y_{j},z_{k-1/2})}{\Delta{z}} \,,
$$
\begin{equation}\label{eq-Maxwell-num3}
    \frac{B_y^{n+1/2}(x_{i+1/2},y_{j},z_{k+1/2}) - B_y^{n-1/2}(x_{i+1/2},y_{j},z_{k+1/2})}{\Delta{t}} =
\end{equation}
$$
= - \frac{E_x^{n}(x_{i+1/2},y_{j},z_{k+1}) - E_x^{n}(x_{i+1/2},y_{j},z_{k})}{\Delta{z}} +
$$
$$
+\frac{E_z^{n}(x_{i+1},y_{j},z_{k+1/2}) - E_z^{n}(x_i,y_{j},z_{k+1/2})}{\Delta{x}} \,.
$$
The rest of equations with ${\partial B_z} /{\partial t}$, $\partial D_y /\partial t $, $\partial D_z / \partial t$ we approximate by similar finite-difference expressions. And we impose the standard constraints on $\Delta t$, $\Delta x$, $\Delta y $, and $\Delta z $ to ensure the stability of an explicit numerical scheme.
\begin{equation}\label{eq-condit-stability}
    \sqrt{ \left(\Delta x\right)^2 + \left(\Delta y\right)^2 + \left(\Delta z\right)^2} > c_{\max} \Delta{t} \,,
\end{equation}
where $c_{\ max}$ is the maximum speed of light $c_{v}=c / \sqrt{\varepsilon \mu}$ in the volume $V_0$ (it is convenient to take as $c_{\max}$ the speed of light in a vacuum).

\section{Results of simulation modeling}
\hspace{0.7 cm}

\begin{table}[!h]
\centering
\renewcommand{\tablename}{Table.}
\caption{Array $\beta_{ij}$ for the angle $\alpha = 90^{\circ}$}
\label{90}
\begin{tabular}{|c|c|c|c|c|c|c|c|c|c|}
\hline
      & Mod.1 & Mod.2 & Mod.3 & Mod.4 & Mod.5 & Mod.6 & Mod.7 & Mod.8 & Mod.9 \\ \hline
Mod.1 & 0     & 0.4  & 0.49  & 0.44   & 0.35  & 0.45  & 0.5  & 0.41  & 0.37  \\ \hline
Mod.2 & 0.4  & 0     & 0.31   & 0.62  & 0.51  & 0.39  & 0.56  & 0.53  & 0.41  \\ \hline
Mod.3 & 0.49  & 0.31   & 0     & 0.6  & 0.53  & 0.56  & 0.46  & 0.48  & 0.42  \\ \hline
Mod.4 & 0.44   & 0.62  & 0.6  & 0     & 0.49  & 0.54   & 0.39  & 0.52  & 0.51  \\ \hline
Mod.5 & 0.35  & 0.51  & 0.53  & 0.49  & 0     & 0.47  & 0.44  & 0.4  & 0.46  \\ \hline
Mod.6 & 0.45  & 0.39  & 0.56  & 0.54   & 0.47  & 0     & 0.57  & 0.52  & 0.46  \\ \hline
Mod.7 & 0.5  & 0.56  & 0.46  & 0.39  & 0.44  & 0.57  & 0     & 0.4  & 0.47  \\ \hline
Mod.8 & 0.41  & 0.53  & 0.48  & 0.52  & 0.4  & 0.52  & 0.4  & 0     & 0.41  \\ \hline
Mod.9 & 0.37  & 0.41  & 0.42  & 0.51  & 0.46  & 0.46  & 0.47  & 0.41  & 0     \\ \hline
\end{tabular}
\end{table}

\begin{table}[!h]
\centering
\renewcommand{\tablename}{Table.}
\caption{Array $\beta_{ij}$ for the angle $\alpha = 45^{\circ}$}
\label{45}
\begin{tabular}{|c|c|c|c|c|c|c|c|c|c|}
\hline
      & Mod.1 & Mod.2 & Mod.3 & Mod.4 & Mod.5 & Mod.6 & Mod.7 & Mod.8 & Mod.9 \\ \hline
Mod.1 & 0     & 0.44  & 0.53  & 0.42   & 0.37  & 0.53  & 0.45  & 0.47  & 0.33  \\ \hline
Mod.2 & 0.44  & 0     & 0.35   & 0.3  & 0.47  & 0.44  & 0.32  & 0.3  & 0.4  \\ \hline
Mod.3 & 0.53  & 0.35   & 0     & 0.36  & 0.53  & 0.36  & 0.35  & 0.4  & 0.46  \\ \hline
Mod.4 & 0.42   & 0.3  & 0.36  & 0     & 0.47  & 0.34  & 0.41  & 0.36  & 0.53  \\ \hline
Mod.5 & 0.37  & 0.47  & 0.53  & 0.47  & 0     & 0.52  & 0.47  & 0.36  & 0.5  \\ \hline
Mod.6 & 0.53  & 0.44  & 0.36  & 0.34   & 0.52  & 0     & 0.3  & 0.47  & 0.49  \\ \hline
Mod.7 & 0.45  & 0.32  & 0.35  & 0.41  & 0.47  & 0.3  & 0     & 0.42  & 0.44  \\ \hline
Mod.8 & 0.47  & 0.3  & 0.4  & 0.36  & 0.36  & 0.47  & 0.42  & 0     & 0.48  \\ \hline
Mod.9 & 0.33  & 0.4  & 0.46  & 0.53  & 0.5  & 0.49  & 0.44  & 0.48  & 0     \\ \hline
\end{tabular}
\end{table}

The processing of results of the Maxwell equations numerical solution requires a transition to a grid in the spherical coordinate system.

Tables 2, 3 show the results of our calculations for two antenna positions on the breast surface. As we see, the relative variations in the electric field power can depend quite strongly on the spatial structure of the biotissue, and it can lead to significant individual deviations of the internal temperature in the model.

\section*{Conclusions}
\hspace{0.7 cm}

We create the software for numerical calculation of electric field in a mammary gland. It takes into account breast's multicomponence and heterogeneity in the distribution of physical parameters. Our model of the power distribution of the radiation field in the $P_d$ inside the mammary gland shows strong relative variations, depending on the small-scale internal structure of the tissue. We calculate RMS deviations of the $P_d$ for 9 different models of the internal structure. They are within $30\div50\%$. This result indicates the need to take into account the small-scale internal structure of the biotissue when processing and analyzing RTM measurement data.

\vspace {5mm} {\it This work was supported by the Russian Foundation for Basic Research and the Administration of the Volgograd Region, project number 15-47-02642 r\_a. }

\begin{biblio}
\bibitem{Barett} Barett A.H., Myers P.C., Sadowsky N.L. Microwave Thermography in the Detection of Breast Canser. {\it American Journal of Roentgenology}, 1980, vol.~34(2), pp.~365--368.
\bibitem{Akki-Arunachalam-2014}  Akki R.S., Arunachalam K. Breast tissue phantoms to assist compression study for cancer detection using microwave radiometry. {\it 36th Annual International Conference of the IEEE Engineering in Medicine and Biology Society}. 2014, pp.~1119--1122.
doi:~10.1109/EMBC.2014.6943791
\bibitem{losev} Losev A.G., Khoperskov A.V., Astakhov A.S., Suleymanova Kh.M. Problems of Measurement and Modeling of Thermal and Radiation Fields in Biological Tissues: Analysis of Microwave Thermometry Data. {\it Science Journal of Volgograd State University. Mathematics. Physics}, 2015, no.\,6 (31), pp.\,98--142. doi:~10.15688/jvolsu1.2015.6.3
\bibitem{novochadov} Novochadov V.V., Shiroky A.A., Khopеrskov A.V., Losev A.G. Comparative modeling the thermal transfer in tissues with volume pathological focuses and tissue engineering constructs: a pilot study. {\it European Journal of Molecular Biotechnology}  2016, vol.\,14, no\,4. pp.\,125--138.
doi:~10.13187/ejmb.2016.14.125
\bibitem{vesnin} Vesnin S.G., Sedakin K.M. Development of Antenna-Applicator Series for Tissue Temperature Non-Invasive Measurement of a Human. {\it Engineering Journal: Science and Innovation}, 2012, no.\,11, pp.\,1--18.
\bibitem{Carr} Carr K.L. Microwave Radiometry: its Importance to the Detection of Cancer. {\it IEEE MTT}, 1989, vol.~37, pp.~12--24.
\bibitem{Foster} Foster K.R., Cheever E.A. Microwave radiometry in biomedicine: a reappraisal. {\it Bioelectromagnetics}, 1992, vol.~13 (6), pp.~567--579.
\bibitem{polyakov} Polyakov  M.V., Khoperskov A.V. Mathematical modeling of radiation fields in biological tissues: the definition of the brightness temperature for the diagnosis. {\it Science Journal of Volgograd State University. Mathematics. Physics}, 2016, no.\,5 (36), pp.\,73--84.
doi:~10.15688/jvolsu1.2016.5.7
\bibitem{Gonzalez} Gonzalez F.J. Thermal simulation of breast tumors. {\it Revista mexicana de fisica}, 2007, vol.~53, pp.~323--326.
\bibitem{Bardati} Bardati F.,~Iudicello S. Modeling the Visibility of Breast Malignancy by a Microwave Radiometer. {\it Biomed. Engineering}, 2008, vol.~55, pp.~214--221. doi:~10.1109/TBME.2007.899354


\end{biblio}

\vspace{5mm}
{\it Maxim Valentinovich Polyakov, Student, Department of Information Systems and Computer Modeling, VolSU, Volgograd, Russia, infomod@volsu.ru}

\vspace{5mm}
{\it Alexander Valentinovich Khoperskov, DSc(Math), Professor, Department
of Information Systems and Computer Modeling, VolSU, Volgograd, Russia,
khoperskov@volsu.ru}

\vspace{5mm}
{\it Andrey Vladimirovich Svetlov, PhD(Math), Assistant professor, Department
of Mathematical Analysis and Function Theory, VolSU, Volgograd, Russia,
andrew.svetlov@volsu.ru}

\begin{flushright}
{ \it Received {\rm April ??, 2017}}
\end{flushright}

{\phantom{eee} \hrule \vskip 4 pt}
\begin{flushleft}
\textbf{УДК 51.76} 
\end{flushleft}
\author{М.В. Поляков, А.В. Хоперсков, А.В. Светлов}
\title{ИМИТАЦИОННОЕ МОДЕЛИРОВАНИЕ РАДИАЦИОННОГО ПОЛЯ В БИОТКАНИ МОЛОЧНОЙ ЖЕЛЕЗЫ}
\maketitle{}
\begin{abstract} \begin{tabular}{p{0mm}p{139mm}}
&\noindent {\footnotesize \qquad Построена математическая модель распределения микроволнового электрического поля в неоднородной биоткани молочной железы для исследования эффективности метода медицинской диагностики на основе микроволновой термометрии. Проведено численное моделирование электромагнитного поля в молочной железе для различных наборов пространственной структуры биоткани. Мелкомасштабная структура обусловлена сложной комбинацией нескольких компонент: кровотоки, жировая ткань, мышечная ткань, молочные дольки, кожа. Варьируя пространственную структуру биоткани, мы оценили влияние этого фактора на распределение электромагнитного поля в объеме молочной железы, которое лежит в основе измерений внутренней температуры методом РТМ.

\qquad\keywordsrus{математическое моделирование, численные методы, биологические ткани, онкология, микроволновая радиотермометрия, перенос тепла, радиационные поля.}}
\end{tabular}\end{abstract}

\begin{biblio_rus}
\bibitem{Barett} Barett, A.H. Microwave Thermography in the Detection of Breast Canser / A.H. Barett, P.C. Myers, N.L. Sadowsky //  American Journal of Roentgenology. --- 1980. --- Vol.~34(2). --- P.~365--368.
\bibitem{Akki-Arunachalam-2014}  Akki, R.S. Breast tissue phantoms to assist compression study for cancer detection using microwave radiometry / R.S. Akki, K. Arunachalam //  36th Annual International Conference of the IEEE Engineering in Medicine and Biology Society. --- 2014. --- P.~1119--1122.
doi:~10.1109/EMBC.2014.6943791
\bibitem{losev}
Лосев, А.Г. Проблемы измерения и моделирования тепловых и радиационных полей в биотканях: анализ данных микроволновой термометрии / А.Г. Лосев,  А.В. Хоперсков, А.С. Астахов, Х.М. Сулейманова  // Вестник Волгоградского государственного университета. Серия 1: Математика. Физика. --- 2015. --- N 6 (31). --- С.\,31--71.
 doi:~10.15688/jvolsu1.2015.6.3
\bibitem{novochadov} Novochadov, V.V. Comparative modeling the thermal transfer in tissues with volume pathological focuses and tissue engineering constructs: a pilot study / V.V. Novochadov, A.A. Shiroky, A.V. Khopеrskov, A.G. Losev //  European Journal of Molecular Biotechnology. ---  2016. --- Vol.\,14. --- N\,4. --- P.\,125--138.
doi:~10.13187/ejmb.2016.14.125
\bibitem{vesnin}
Веснин, С.Г. Разработка серии антенн-аппликаторов для неинвазивного измерения температуры тканей организма человека при различных патологиях / С.Г. Веснин, К.М. Седакин // Инженерный журнал: наука и инновации. --- 2012. --- N 11. --- C. 1--18.
\bibitem{Carr} Carr, K.L. Microwave Radiometry: its Importance to the Detection of Cancer / K.L. Carr //  IEEE MTT. --- 1989. --- Vol.~37. --- P.~12--24.
\bibitem{Foster} Foster, K.R. Microwave radiometry in biomedicine: a reappraisal / K.R. Foster, E.A. Cheever  // { Bioelectromagnetics}. --- 1992. --- Vol.~13 (6). --- P.~567--579.
\bibitem{polyakov}
Поляков, М.В. Математическое моделирование пространственного распределения радиационного поля в биоткани: определение яркостной температуры для диагностики / М.В. Поляков, А.В. Хоперсков  // Вестник Волгоградского государственного университета. Серия 1: Математика. Физика. --- 2016. --- N 5 (36). --- С. 73--84.
doi:~10.15688/jvolsu1.2016.5.7
\bibitem{Gonzalez} Gonzalez, F.J. Thermal simulation of breast tumors / F.J. Gonzalez // { Revista mexicana de fisica}. --- 2007. --- Vol.~53. --- P.~323--326.
\bibitem{Bardati} Bardati, F. Modeling the Visibility of Breast Malignancy by a Microwave Radiometer /  F. Bardati, S. Iudicello //  { Biomed. Engineering}. --- 2008. --- Vol.~55. --- P.~214--221. doi:~10.1109/TBME.2007.899354

\end{biblio_rus}

\vspace{5mm}
{\it Поляков Максим Валентинович, кафедра Информационных систем и компьютерного моделирования, Волгоградский государственный университет,  infomod@volsu.ru}

{\it Хоперсков Александр Валентинович, д.ф.-м.н., профессор, кафедра Информационных систем и компьютерного моделирования, Волгоградский государственный университет,  infomod@volsu.ru}

{\it Светлов Андрей Владимирович, к.ф.-м.н., доцент, кафедра Математического анализа и теории функций, Волгоградский государственный университет, andrew.svetlov@volsu.ru}

\begin{flushright}
{ \it Поступила в редакцию {\rm ?? ??? 2017 г.}}
\end{flushright}

\end{document}